\documentstyle[aps]{revtex}
\def\be{\begin{equation}}
\def\ee{\end{equation}}
\def\bea{\begin{eqnarray}}
\def\eea{\end{eqnarray}}
\begin{document}  

\title{Doublet-Triplet Splitting in Supersymmetric SU(6) By Missing VEV 
Mechanism}

\author{ Z. Chacko and Rabindra N. Mohapatra}
\address{\it Department of Physics, University of Maryland,
College Park, MD 20742, USA}

\maketitle

\begin{abstract}
We present a realistic supersymmetric SU(6) model which implements
doublet-triplet splitting by the missing vev mechanism. The model 
makes use of only the simplest representations, requires no
fine tuning of parameters and maintains coupling constant unification 
as a prediction. Fermion masses also emerge in a very straightforward 
manner.
 This is the first time that the missing vev mechanism 
has been realized in the context of SU(6).
\end{abstract}

\hskip 4cm UMD-PP-99-019~~~~e-mail address: rmohapat@physics.umd.edu
 
\section{Introduction}\hspace{0.5cm} 
Satisfactory resolution of the Higgs mass problem as well as the 
radiative origin of electroweak symmetry breaking in the 
supersymmetric extension of the standard model (MSSM) has rightly focussed 
the interest of the particle 
theory community in understanding the high energy origin of the MSSM.
The unification of the gauge couplings for the MSSM particle content
gives rise to the natural speculation that the high energy theory may
indeed be a supersymmetric grand unified theory\cite{1,2} based on some 
simple group.
Some popular groups which have been investigated are: SU(5), SO(10),
$E_6$ and SU(6). In this paper, we will concentrate on the SU(6) model.

A key problem of all SUSY GUTs is how to split the weak MSSM doublets 
from the color triplet fields that accompany them as part of the 
representation of the
GUT symmetry. This is an essential aspect of SUSY GUTs since both 
coupling constant unification and suppression of proton decay require that
the MSSM doublets $H_u$ and $H_d$ be at the weak scale whereas the 
triplets which mediate proton decay must have GUT scale mass. This is the 
famous doublet-triplet splitting problem (DTS).

In the simplest grand unifying group SU(5) natural implementation
of doublet triplet splitting seems to require the use of the 
relatively large representations $\bf{50}, \overline{\bf{50}}$ and {\bf{75}}.
The search for a simpler solution has lead various authors to consider the
extension of the gauge symmetry to SU(6) which allows for more 
possibilities. Some ideas that have received
considerable attention are: (i) the ``GIFT'' mechanism, where the light 
Higgs doublets emerge as the pseudo-Goldstone bosons of a broken global 
symmetry of the superpotential\cite{bere} and (ii) the sliding singlet 
mechanism\cite{wit}, where the desired vev pattern follows automatically
from the conditions of the supersymmetric minima condition. 
Although the ``GIFT'' mechanism is very interesting as an idea, it leads to 
complications while trying to construct realistic models, 
specially in obtaining masses for the quarks and leptons. Future 
developments in implementing this elegant idea are awaited. 
On the other hand, the sliding singlet mechanism, which is perhaps the most
straightforward of all doublet-triplet splitting mechanisms has its own 
shortcomings. After this was first proposed in the context of SU(5) model, 
it was realized that it
runs into severe difficulties due to radiative corrections which actually 
lift the MSSM doublet masses to an intermediate scale\cite{wit,haba}. It 
was shown by 
Sen\cite{sen} that if instead one considers an embedding of the SU(5) model 
into the
SU(6) group, then the problems associated with radiative instability of 
doublet-triplet splitting can be cured. The observation of Sen also made 
it clear that the SU(6) model may have certain technical advantages over 
other GUT groups as far as understanding the DTS problem goes. This
version of the SU(6) model however is ruled out by the present data on 
$sin^2\theta_W$. An interesting variation of Sen's idea which cures this 
difficulty has been discussed in a recent paper\cite{barr}. 

In this paper we show that the missing vev mechanism which has been 
successfully implemented in supersymmetric SO(10) can also be realized
in a simple manner in SU(6), providing yet another way to solve 
the DTS problem in these models. We present a realistic model which makes
use of this idea to implement DTS without finetuning of any parameters
and which maintains coupling constant unification as a prediction.
The Higgs sector of the model is quite economical and
makes use of only the simplest representations; namely
two adjoints ({\bf 35}s), two ${\bf 6 +\bar{\bf{6}}}$ pairs and a singlet 
field.
Fermion masses also emerge in a more straightforward manner than in earlier 
schemes.

The basic idea behind our model is to obtain a pattern of vevs for one of 
the adjoints $S$ of the form:
\begin{eqnarray}
<S>= const.\left(\begin{array}{cccccc}
-3 & & & & & \\
 & 1 & & & & \\
 & & 1 & & & \\
 & & & 1 & & \\
 & & & & 0 & \\
 & & & & & 0  \end{array} \right)
\end{eqnarray}
where the lower $5\times  5$ block corresponds to the fields which 
transform under the SU(5) which contains the standard model. Then the
two $SU(2)_L$ doublets in one of the ${\bf 6 + \bar{6}}$ pairs (denoted 
by $h +\bar{h}$) remain massless and become the MSSM doublets via a 
coupling of the form $S h\bar{h}$ without any finetuning of parameters.
Since the adjoint $S$ does not break $SU(6)$ all the way down to the 
standard model (it leaves an extra U(1)), we will need additional Higgs
fields, which are the other the ${6+\bar{6}}$ pair. It is important to 
point out that a similar scheme is not feasible in the case of the SU(5) 
model since the adjoint 
must be traceless and the pattern of symmetry breaking implies that there be 
only two different vevs along the diagonal.

\section{Symmetry breaking and Doublet-Triplet Splitting via Missing Vev}

We now describe the Higgs sector of the model which breaks SU(6) down 
to the standard model while keeping supersymmetry intact and show how
in this process doublet-triplet splitting is realized. The 
Higgs sector of the model as already mentioned consists of two adjoints 
$\Sigma$, $\bar{\Sigma}$ ({\bf 35}-dim.), two pairs of ${\bf 6+\bar{6}}$
denoted by ${\bf h +\bar{h}}$ and ${\bf H+\bar{H}}$ and a singlet $Y$. We
will assume the MSSM doublets to emerge from ${\bf h +\bar{h}}$ pair. We 
choose the Higgs superpotential to have the form:
\begin{eqnarray}
W_{Higgs} = Y(\Sigma^2-\bar{\Sigma}^2)+ (\Sigma^3 
+\bar{\Sigma}^3)\nonumber \\+H(\Sigma +\bar{\Sigma})\bar{H} +MH\bar{H}
+h(\Sigma +\bar{\Sigma})\bar{h} 
\end{eqnarray}
This superpotential is invariant under the discrete symmetry:
\begin{eqnarray}\Sigma \leftrightarrow \bar{\Sigma}; Y\leftrightarrow -Y;
H\leftrightarrow -H; \bar{H}\leftrightarrow -\bar{H}
\end{eqnarray}
Fields not included in the above equation are invariant under the 
discrete symmetry. Freedom to rescale the fields has been used to set all 
the dimensionless
coupling parameters in the superpotential to one.

 We are interested in the supersymmetric vacua of the following form:
\begin{eqnarray}
<H>=<\bar{H}>=\left(\begin{array}{c} v \\ 0 \\0 \\0 \\0 \\ 0  
\end{array}\right) \nonumber\\
<\Sigma> =\left( \begin{array}{ccc}
a_1 &  &  \\
 & a_2 {\bf I_3} & \\
  &   & a_5{\bf I_2}\end{array}\right)
\nonumber\\
<\bar{\Sigma}> =\left( 
\begin{array}{ccc}                                           
 \bar{a}_1 &  &  \\                                                                   
 & \bar{a}_2 {\bf I_3} & \\                                                          
  &   &\bar{a}_5{\bf I_2}\end{array}\right)                                         
\end{eqnarray}                                                          
Here ${\bf I_n}$ stands for an $n\times n$ unit matrix. These vev's break 
SU(6) down to $SU(3)\times SU(2)\times U(1)$. Now let us note that if the 
vacuum conditions are such that $a_5=-\bar{a}_5$, then $\Sigma 
+\bar{\Sigma}$ will have the vev form of $<S>$ in Eq. (1) and the last 
term in the superpotential in Eq.(2) then implies that the weak doublets in
$h+\bar{h}$ are massless after the SU(6) breaking. We now show that vacuum
conditions for maintaining supersymmetry indeed imply that 
$a_5=-\bar{a}_5$.  

The equations of motion for the various fields are:
\begin{eqnarray}
M+ (a_1 +\bar{a}_1)=0\nonumber \\
\Sigma_i a^2_i = \Sigma_i \bar{a}^2_i \nonumber \\
2ya_i + \alpha + 3 a^2_i + v^2\delta^i_1 =0 \nonumber\\
-2 y a_i +\bar{\alpha} + 3\bar{a}^2_i + v^2\delta^i_1=0
\end{eqnarray}
where $\alpha$ and $\bar{\alpha}$ are the Lagrange multipliers that are 
needed to guarantee the tracelessness of  the $\Sigma$ and $\bar{\Sigma}$
vevs. Summing the last two equations in the Eq. (5) and subtracting and
remembering that $\Sigma_i a_i=\Sigma_i \bar{a}_i=0$, we get 
$\alpha=\bar{\alpha}$. Then subtracting the same two last equations of 
Eq.(5), we get:
\begin{eqnarray}
(a_i +\bar{a}_i)(a_i-\bar{a}_i +\frac{2}{3}y)=0
\end{eqnarray}
This implies that either 
\begin{eqnarray}
a_i=-\bar{a}_i \nonumber \\
or ~~~~~~~\bar{a}_i=a_i +\frac{2}{3} y
\end{eqnarray}
We are interested in the vacuum where the first of the equations in 
Eq.(7) is satisfied by $a_5$ whereas $a_{1,2}$ satisfy the second of the 
Eq. (7). Then $\Sigma_i a_i=\Sigma_i\bar{a}_i=0$ implies that
\begin{eqnarray}
a_5~=~\frac{2}{3}y
\end{eqnarray}
The equation of motion for $\Sigma$ in Eq.(5) is quadratic which then implies 
that
\begin{eqnarray}
a_2 + a_5 = -\frac{2}{3} y
\end{eqnarray}
leading to $a_2=-\frac{4}{3} y$. The trace condition then implies that
$a_1= \frac{8}{3} y$. The first equation in Eq.(5) then determines 
$y=-M/6$ and $v^2=-\frac{2}{3}M^2$. Thus all the vevs are determined in 
terms of the mass parameter $M$ of the superpotential. The vevs of
the {\bf 35}s are given by:
\begin{eqnarray}
<\Sigma> = -\frac{M}{9}\left(\begin{array}{cccccc}
4 & & & & & \\
 & -2 & & & & \\
 & & -2 & & & \\
& & & -2 & & \\
 & & & &1 & \\
 & & & & &1 \end{array}\right) \nonumber \\
<\bar{\Sigma}> = -\frac{M}{9}\left(\begin{array}{cccccc}
5 & & & & & \\
 & -1 & & & & \\
 & & -1 & & & \\
 & & & -1 & & \\
 & & & & -1 & \\
& & & & & -1 \end{array}\right)
\end{eqnarray}
 	It is now obvious that $<\Sigma +\bar{\Sigma}>$ has the desired form
and the doublets in $h+\bar{h}$ are now massless whereas the triplets in 
those multiplets have GUT scale mass thus solving the DTS problem of 
SU(6) in a natural manner.

\section{Fermion masses in SU(6)}

It is relatively straightforward to incorporate fermions into the model 
and get desired masses for them. Following a procedure identical to that 
in Ref.\cite{markus}, we add the fields $\bar{P}_{1j,2j}$ belonging to $\bf 
{\bar{6}}$, $N_j$ belonging to ${\bf 15}$ dim. representation, an extra
${\bf 15 +\bar{15}}$ pair denoted by $Z+\bar{Z}$ and a pair of singlets
$T$ and $\bar{T}$. (Here $j$ is the 
generation index.) The known fermions live in $N$ and $\bar{P}_2$.
The part of the superpotential responsible for fermion 
masses is given by
\begin{eqnarray}
W_{fermion} = \lambda^{jk} N_j \bar{H}\bar{P}_{1k} +Y^{jk}_d N_j \bar{P}_{2k}
\bar{h} + TH\bar{P}_{1} + \bar{T}H\bar{P}_{2}\nonumber \\
+Y^{jk}_u N_j N_k Z +M' Z\bar{Z} +\bar{Z}Hh
\end{eqnarray}
The discrete symmetry of Eq. (2) can be extended to this part of the 
superpotential as follows:
\begin{eqnarray}
\bar{P}_1\rightarrow i \bar{P}_1; \bar{P}_2 \rightarrow -i\bar{P}_2:
N\rightarrow i N;\nonumber \\
Z, \bar{Z}\rightarrow -Z, -\bar{Z}; T\rightarrow iT; 
\bar{T}\rightarrow -i\bar{T}
\end{eqnarray} 
The role of the various terms in the superpotential $W_{fermion}$ are as 
follows. The term $N\bar{H}\bar{P}$ along with the terms containing the 
singlets decouple one of the ${\bf 
\bar{6}}$
fields and the unwanted fields in the {\bf 15} to the GUT scale, leaving 
three light generations as desired. The $Y_d$ term gives mass to the down 
type quarks and the charged leptons. The $M'$ and the $\bar{Z}Hh$ terms 
combine to have the MSSM $H_u$ as an admixture of the $Z$ and the 
$\bar{h}$ field. Thus after SUSY breaking the up-vev will reside partly 
in both these fields. Now it is easy to see that the $NNZ$ coupling give 
masses to the up type quarks. Upto now the fermion masses are exactly as 
in the case of SUSY SU(5) and therefore are plagued by the ``bad'' relation
$m_e/m_{\mu}=m_d/m_s$. This can however be cured by a nonrenormalizable
term in the superpotential of the form:
\begin{eqnarray}
W_{nr}= \frac{\lambda'_{jk}}{M_{Pl}} N_j (\Sigma 
+\bar{\Sigma})\bar{P}_{2k}\bar{h}
\end{eqnarray}

Let us now turn our attention to the prediction for proton lifetime in 
our model. As the model has been constructed so far, proton lifetime is same
as in the minimal SUSY SU(5) model\cite{nath} i.e. the Higgsinos whose
exchange leads to proton instability must be heavier than the GUT scale
for proton lifetime to be compatible with present experiments. It may 
therefore be useful to see whether by simple modification of the model, 
one can have suppression of proton decay. Indeed it turns out as we show 
now that a weak suppression is not hard to achieve in our model without
effecting its other attractive features. All we have to do is to 
supplement the Higgs spectrum by adding another pair of ${\bf 6+\bar{6}}$
pair (denoted by ${\bf h'+ \bar{h}'}$) and replacing the 
$h(\Sigma+\bar{\Sigma})\bar{h}$ term in the $W_{Higgs}$ to the following 
form:
\begin{eqnarray}
W'_{Higgs} = h(\Sigma +\bar{\Sigma})\bar{h}' + h'(\Sigma 
+\bar{\Sigma})\bar{h}+ M'' h'\bar{h}'
\end{eqnarray}
The proton decay amplitude now has an extra suppression factor proportional
to $M''/M_U$ where $M_U$ is the GUT scale. If we keep this factor to be 
somewhat less than one then the proton lifetime will be suppressed by a 
decent factor   
over the SUSY SU(5) prediction. There is of course a price for this i.e.
it will generate some light fields just below the scale $M_U$. This will 
slightly enhance the threshold effects with a marginal effect on coupling 
constant unification

It is conceivable that the theory as written down or some straightforward 
generalization thereof is the most general possible that is consistent with
some set of discrete symmetries. However since the nonrenormalization 
theorem of supersymmetry protects the superpotential from radiative
corrections this is not necessary and we do not pursue the matter further 
here.
                                                                         
In conclusion, we have demonstrated how the missing vev mechanism to 
realize doublet-triplet splitting can be  implemented in the supersymmetric 
SU(6) grand unification model without any fine tuning of parameters.
This is the first time such a model has been constructed. We have also
shown how a realistic fermion spectrum and weak suppression of proton decay
can be achieved consistently within this scenario.

 This work is supported by the National Science Foundation 
under grant no. PHY-9802551.

\end{document}